\documentclass[apjl]{emulateapj}
\usepackage{apjfonts}
\usepackage{color}
\usepackage{amsmath}
\usepackage{textcomp}
\usepackage{url}
\usepackage{graphicx, subfigure}

\newcommand{\kms}{\hbox{km~s$^{-1}$}}
\newcommand{\Mjup}{$M_{\mathrm{Jup}}$\ }

\slugcomment{Accepted for publication in ApJ Letters.}

\shorttitle{The Coolest BD candidate in TWA}
\shortauthors{Gagn\'e et al.}

\begin{document}

\title{THE COOLEST ISOLATED BROWN DWARF CANDIDATE MEMBER OF TWA}

\author{Jonathan Gagn\'e\altaffilmark{1}\thanks{jonathan.gagne@astro.umontreal.ca},\,Jacqueline K. Faherty\altaffilmark{2,3,4}\thanks{jfaherty17@gmail.com},\, Kelle Cruz\altaffilmark{5,6},\, David Lafreni\`ere\altaffilmark{1},\, Ren\'e Doyon\altaffilmark{1},\, Lison Malo\altaffilmark{1},\, \'Etienne Artigau\altaffilmark{1}.}

\affil{\altaffilmark{1}D\'epartement de Physique and Observatoire du Mont-M\'egantic, Universit\'e de Montr\'eal, C.P. 6128 Succ. Centre-ville, Montr\'eal, Qc H3C 3J7, Canada \and \altaffilmark{2}Department of Terrestrial Magnetism, Carnegie Institution of Washington, Washington, DC 20015, USA\\
\and \altaffilmark{3}Departamento de Astronom\'ia, Universidad de Chile, Cerro Cal\'an, Las Condes, Chile\\
\and \altaffilmark{4}Hubble Fellow\\ \and \altaffilmark{5}Department of Astrophysics, American Museum of Natural History, Central Park West at 79th Street, New York, NY 10034\\ \and \altaffilmark{6}Department of Physics \& Astronomy, Hunter College, 695 Park Avenue, New York, NY 10065, USA.}

\begin{abstract}
We present two new late-type brown dwarf candidate members of the TW~Hydrae association (TWA)~: 2MASS~J12074836-3900043 and 2MASS~J12474428-3816464, which were found as part of the BANYAN all-sky survey (BASS) for brown dwarf members to nearby young associations. We obtained near-infrared (NIR) spectroscopy for both objects (NIR spectral types are respectively L1 and M9), as well as optical spectroscopy for J1207-3900 (optical spectral type is L0$\gamma$), and show that both display clear signs of low-gravity, and thus youth. We use the BANYAN~II Bayesian inference tool to show that both objects are candidate members to TWA with a very low probability of being field contaminants, although the kinematics of J1247-3816 seem slightly at odds with that of other TWA members. J1207-3900 is currently the latest-type and the only isolated L-type candidate member of TWA. Measuring the distance and radial velocity of both objects is still required to claim them as bona fide members. Such late-type objects are predicted to have masses down to 11~\textendash~15~\Mjup at the age of TWA, which makes them compelling targets to study atmospheric properties in a regime similar to that of currently known imaged extrasolar planets.
\end{abstract}


\section{Introduction}

The known population of brown dwarfs (BDs) has significantly increased in the last decades due to all-sky near-infrared (NIR) surveys such as 2MASS and \emph{WISE} (\citealp{2006AJ....131.1163S}, \citealp{2010AJ....140.1868W}). The acumulation of a large number of BDs allowed for a better understanding of the underlying physics in their atmospheres, which went along with the development of increasingly more realistic atmosphere models (\citealp{2003A&A...402..701B}, \citealp{2008ApJ...689.1327S}, \citealp{2012ApJ...756..172M}, \citealp{2013MSAIS..24..128A}) and empirical spectral classification schemes (\citealp{1991ApJS...77..417K}, \citealp{2005ApJ...623.1115C}, \citealp{2006ApJ...637.1067B}, \citealp{2009AJ....137.3345C}, \citealp{2013ApJ...772...79A}). These tools allowed in turn the identification of peculiar BDs, most of which are now recognized as having atypical metallicity or surface gravity. \\
Low surface gravity BDs are thought to be younger than several hundred million years since they have not yet reached their equilibrium radii \citep{2001RvMP...73..719B}. The youngest and latest-type of these objects are believed to have cool, low-pressure atmospheres similar to those of currently known imaged gaseous giant exoplanets, but only a few of those are known in the solar neighborhood (e.g. 2MASS~J03552337+1133437; \citealp{2013AJ....145....2F}; PSO J318.5338-22.8603; \citealp{2013ApJ...777L..20L}; CFBDSIR 2149-0403; \citealp{2012A&A...548A..26D}). Hence, atmosphere models for such physical conditions are still subject to poor empirical constraints (e.g. the behavior of dust in these low-pressure environments). While the luminosity, equivalent width of atomic lines, and shape of the continuum can be used to identify young brown dwarfs, there is no evidence yet that those can be used to narrowly constrain ages \citep{2013ApJ...772...79A}. Therefore, assembling an an age-calibrated sample identified by kinematics could potentially help addressing this in an empirical way. Given their relative proximity, nearby, young associations (NYAs) such as TW Hydrae (TWA; \citealp{2004ARA&A..42..685Z}) are perfect test benches for such empirical calibrations. The search for late-type objects in NYAs has been the subject of many efforts (\citealp{2004ARA&A..42..685Z}; \citealp{2007ApJ...669L..97L}; \citealp{2008hsf2.book..757T}; \citealp{2013ApJ...762...88M}), however their late-type ($>$ M5) population is poorly constrained. To address this further, \citeauthor{2014ApJ...783..121G} (\citeyear{2014ApJ...783..121G}; called G2014 hereafter) developed Bayesian Analysis for Nearby Young AssociatioNs II (BANYAN~II), a tool based on \cite{2013ApJ...762...88M} that uses naive Bayesian inference to identify late-type candidate members to such NYAs from their sky position, proper motion and photometry. Using this new tool, our team has initiated the BANYAN all-sky survey (BASS) that generated hundreds of $>$ M5 candidate members to NYAs from the 2MASS and \emph{WISE} surveys, using both catalogues as a baseline for a proper motion measurement. The current status of this project is described in more detail in \cite{2013arXiv1307.1127G}. \\
Here, we present two of the potential latest-type and lowest-mass objects that were identified as candidate members to TWA from this all-sky survey : 2MASS~J12474428-3816464 (M9; called J1247-3816 hereafter) and 2MASS~J12074836-3900043 (L1; called J1207-3900 hereafter), with NIR spectral types M9 and L1, respectively. We present NIR SpeX spectroscopy for the two objects, as well as optical MAGE spectroscopy for J1207-3900 in Section~\ref{sec:spt}. In Section~,\ref{sec:lowg}, we show evidence that both have a low surface gravity, and we use the BANYAN~II tool in Section~\ref{sec:bayes} to show that both objects are likely members of TWA with a small probability of being young field contaminants, but that J1247-3816 seems to display slightly discrepant kinematics.

\section{Spectroscopy}\label{sec:spectro}
\subsection{NIR Spectroscopy}\label{sec:nir}

We have obtained SpeX (\citealp{2003PASP..115..362R}) NIR spectroscopy for J1207-3900 and J1247-3816 at the IRTF telescope on May 10 2013. Observations were obtained under a typical seeing of 0\textquotedbl .6. We used the prism disperser with the 0\textquotedbl .8 slit for both objects, yielding a resolution R $\sim 95$ over 0.8 to 2.5 $\mu$m. Four exposures of 200 seconds for J1207-3900 and 180 seconds for J1247-3816 were sufficient to reach signal-to-noise (S/N) per resolution element $\sim 240$ for both objects. We have subsequently obtained an R $\sim 750$ spectrum for J1207-3900 in the cross-dispersed mode with the 0\textquotedbl .8 slit on May 14 2013 to be able to measure the equivalent width of several atomic lines and better constrain its low-gravity using the approach of \citeauthor{2013ApJ...772...79A} (\citeyear{2013ApJ...772...79A}; see Section \ref{sec:lowg}). Ten exposures of 200 seconds yielded a S/N per resolution element $\sim 65$. Individual exposures were reduced by subtracting dithered sequences along the slit, extracting both traces and correcting for 

\begin{figure*}
	\centering
	\subfigure[2MASS~J12074836-3900043]{\label{fig:NIR_compa}\includegraphics[width=0.99\textwidth]{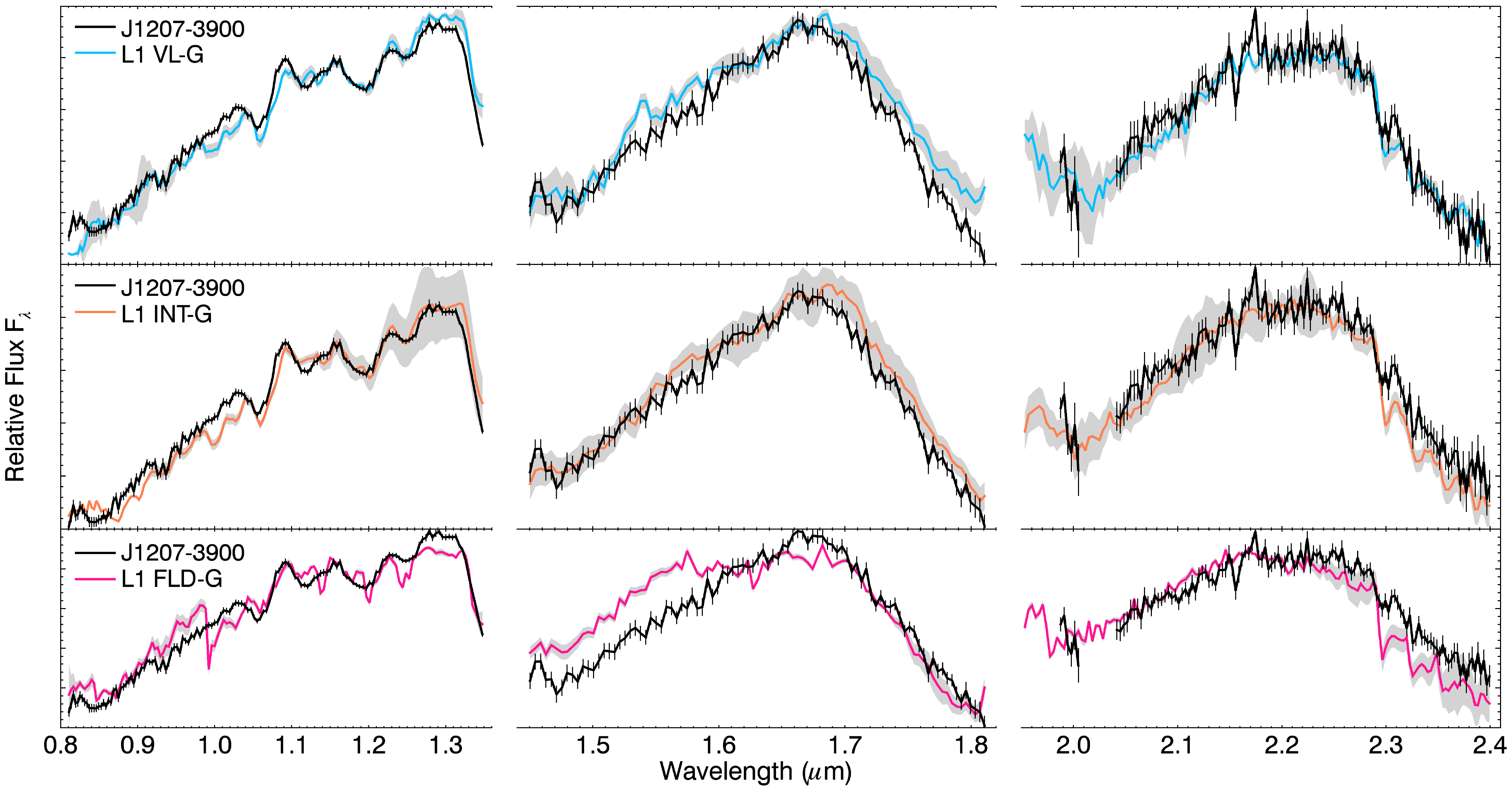}}
	\subfigure[2MASS~J12474428-3816464]{\label{fig:NIR_compb}\includegraphics[width=0.99\textwidth]{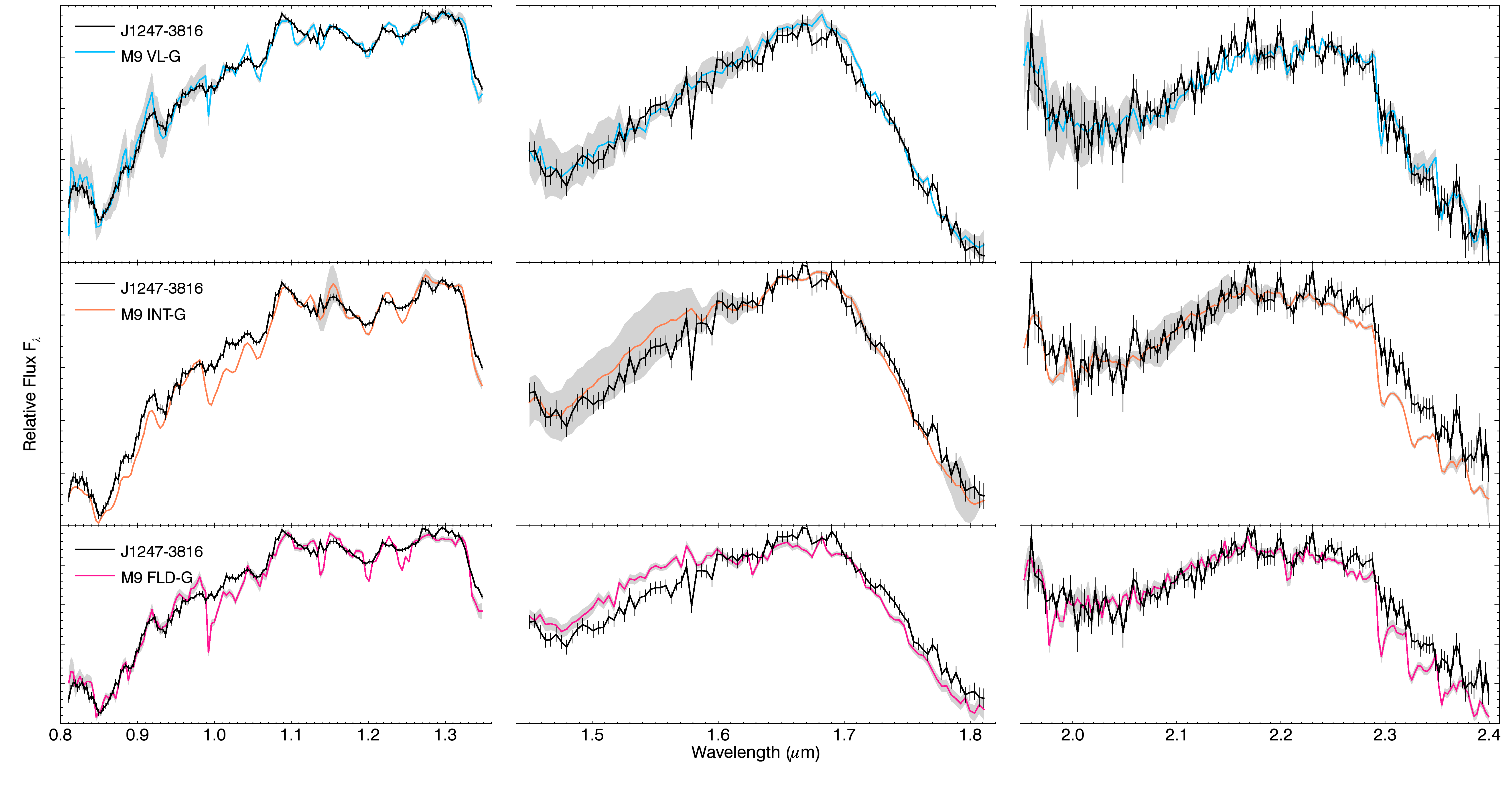}}	
	\caption{Comparison of the NIR spectra of J1207-3900 (a) and J1247-3816 (b; black lines) with L1 and M9 spectroscopic templates (colored lines; as described in Section~\ref{sec:spt}), respectively. Each band was normalized individually. The grey shaded region represents the scatter between individual objects that were used to create the templates and the black vertical lines represent the measurement errors on each bin of the candidates' spectra. In the case of J1247-3816, a better match is clearly achieved with the very low gravity template, whereas J1207-3900 is well fit by both the VL-G and INT-G templates which are quite similar themselves.}
	\label{fig:NIR_comp}
\end{figure*}

\LongTables
\begin{deluxetable}{lcc}
\tablecolumns{6}
\tablecaption{Properties of the New Candidates \label{tab}}
\tablehead{\colhead{Property} & \colhead{J1207-3900} &  \colhead{J1247-3816}}
\startdata
RA & 12:07:48.362 & 12:47:44.290\\
DEC & -39:00:04.40 & -38:16:46.40 \\
$\mu_\alpha$ (mas yr$^{-1}$) & $-57.2 \pm 7.9$ & $-33.2 \pm 7.1$\\
$\mu_\delta$ (mas yr$^{-1}$) & $-24.8 \pm 10.5$ & $-16.6 \pm 9.5$\\
$I$ (\emph{DENIS}) & $\cdots$ & $17.85 \pm 0.16$\\
$J$ (2MASS) & $15.50 \pm 0.06$ & $14.79 \pm 0.03$\\
$H$ (2MASS)& $14.61 \pm 0.05$ & $14.10 \pm 0.04$\\
$K_s$ (2MASS) & $14.04 \pm 0.06$ & $13.57 \pm 0.04$ \\
$W1$ (\emph{WISE}) & $13.63 \pm 0.03$\tablenotemark{a} & $13.11 \pm 0.02$\\
$W2$ (\emph{WISE}) & $13.22 \pm 0.03$ & $12.52 \pm 0.03$ \\
$W3$ (\emph{WISE}) & $>~13.20$ & $10.95 \pm 0.08$\\
$W4$ (\emph{WISE}) & $>~9.20$ & $8.84 \pm 0.29$\tablenotemark{b}\\
\hline
Optical Spectral type & L0$\gamma \pm 0.5$ & $\cdots$\\
NIR Spectral type & L1 $\pm 1$ VL-G & M9 $\pm 0.5$ VL-G\\
$TWA~d_{s}$\tablenotemark{c} (pc) & $60.2 \pm 5.2$ & $63.8 \pm 6.4$\\
$TWA~v_{s}$\tablenotemark{c} (km s$^{-1}$) & $9.7 \pm 1.7$ & $9.6 \pm 1.7$\\
$Field~d_{s}$\tablenotemark{c} (pc) & $63.0 \pm 10.4$ & $55.4 \pm 11.6$\\
$Field~v_{s}$\tablenotemark{c} (km s$^{-1}$) & $6.7 \pm 10.2$ & $4.0 \pm 10.4$
\enddata
\tablenotetext{a}{Possibly contaminated by a diffraction spike.}
\tablenotetext{b}{Possibly contaminated by a nearby source.}
\tablenotetext{c}{Statistical predictions from the BANYAN~II tool. See Section~\ref{sec:bayes} for more information.}
\end{deluxetable}

\noindent telluric absorption with A0-type standards, using the SpeXtool Interactive Data Language (IDL) package (\citealp{2004PASP..116..362C}; \citealp{2003PASP..115..389V}). The NIR spectra for both objects are displayed in Figure~\ref{fig:NIR_comp}.

\subsection{Optical Spectroscopy}\label{sec:opt}

In addition to the NIR spectroscopy described in the previous section, we have obtained optical spectroscopy for J1207-3900 on May 14 2013 with MagE at the Magellan telescope to compare it with standard optical spectra of low-gravity BDs (\citealp{2009AJ....137.3345C}). We used the 0\textquotedbl .7 slit and 2800 seconds of exposure to obtain a R $\sim 5800$ spectrum in the 5500~\textendash~10300~\AA\ range with a S/N per resolution element $\sim 16$. Individual exposures were reduced in a similar manner than described in the previous section by using the MASE IDL package \citep{2009PASP..121.1409B}. The optical spectrum of J1207-3900 is presented in Figure~\ref{fig:OPT_comp}\footnote[1]{All spectra presented here for J1207-3900 and J1247-3816 can be found at \url{www.astro.umontreal.ca/\textasciitilde gagne}.}.

\section{Results and Discussion}\label{sec:discussion}

\subsection{Spectral Classification}\label{sec:spt}

\begin{figure*}
	\begin{center}
		 \includegraphics[width=0.99\textwidth]{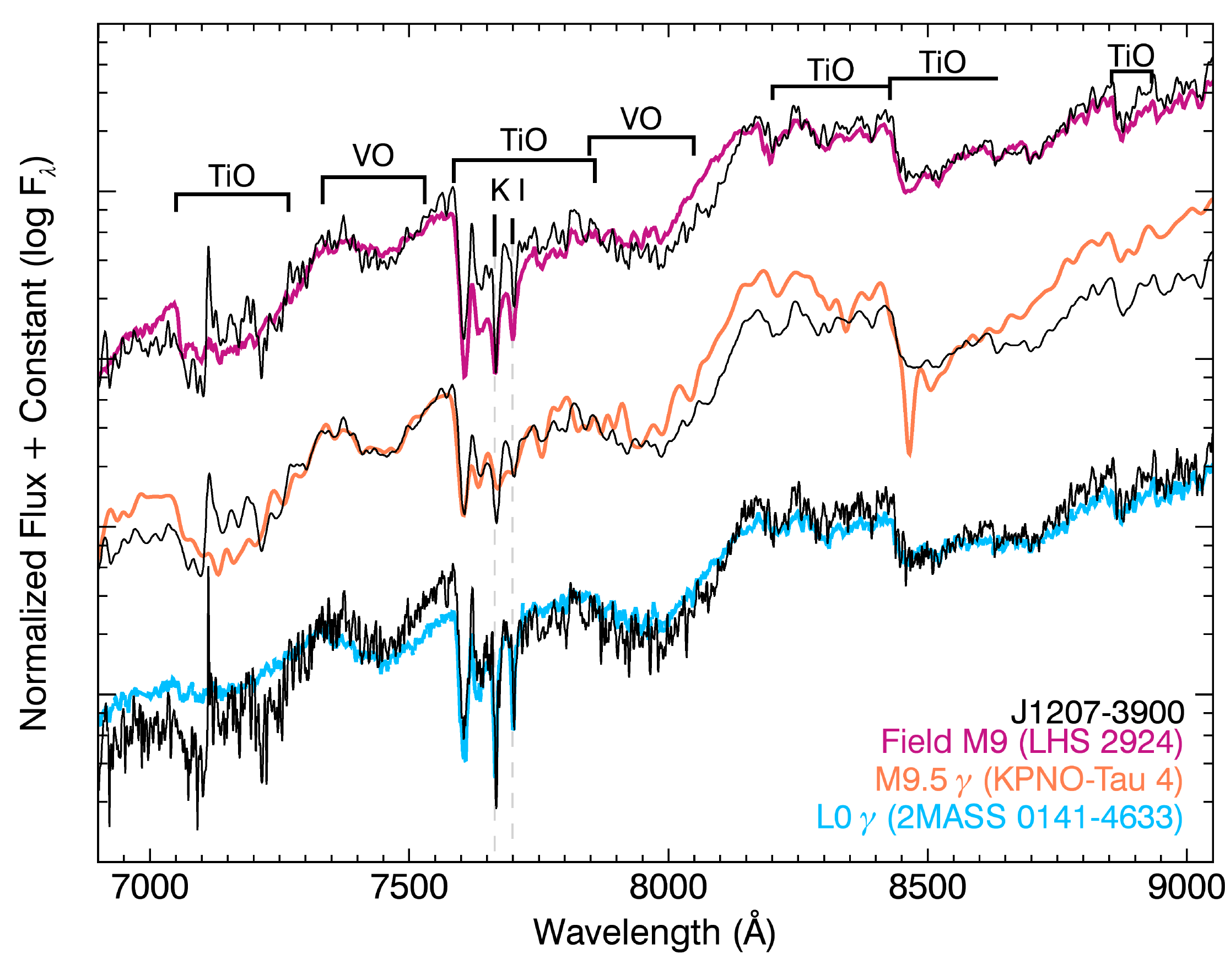}
	\end{center}
	\caption{Optical MAGE spectrum of J1207-3900, compared to known field and young late M standards (\citealp{2003IAUS..211..163S} ; \citealp{2005PASP..117..676R} ; \citealp{2006AJ....131.1007B} ; \citealp{2008AJ....136.1290R}). All spectra were normalized to their median between 7000 and 9000 \AA. It can be seen that the VO band at 7450 \AA\ is deeper than that of field dwarfs in J1207-3900, which is a telltale sign of youth.}
	\label{fig:OPT_comp}
\end{figure*}

We used the method of K. Cruz et al. (in preparation)\footnote[2]{see also the 2012 Cool Stars 17 poster \citealp{DisentanglingLDwar:db}} to median-combine all NIR spectra from \cite{2013ApJ...772...79A} by spectral type to create individual NIR spectroscopic templates for intermediate-gravity (INT-G) objects in the M8~\textendash~L3 range, as well as very low gravity (VL-G) objects in the M6~\textendash~L4 range. We used objects that were classified as having a normal surface gravity, as well as medium-resolution spectra from the SpeX Prism library to build field NIR templates in the M5~\textendash~L9 range. We then assigned spectral types to both objects by visually comparing their spectra band-by-band with those composite spectroscopic templates (see Figure~\ref{fig:NIR_comp}). We find that the M9 VL-G template is clearly the best match to J1247-3816 and that the L1 INT-G and L1 VL-G templates are equally good matches to J1207-3900. We have visually assigned uncertainties of $\pm$ 0.5 and $\pm$ 1 subtypes, respectively. If we restrict our comparison to field NIR standards only, we also find that M9 and L1 spectral types are the best matches, however we would have assigned larger uncertainties to them. We also directly compared the spectra of the two candidates to the latest currently known TWA members and candidates (TWA~28, M8.5 candidate; TWA~26, M9 member; and TWA~29, M9.5 candidate; see \citealp{2005ApJ...634.1385M} and \citealp{2007ApJ...669L..97L}) to confirm our results. Since there are no known L0 \textendash\ L4 objects in TWA, we could only verify that J1207-3900 has redder \emph{J}\textendash\ and \emph{H}\textendash band slopes than TWA 29, which is consistent with it being later-type.\\
We have subsequently compared the MAGE optical spectrum to several field and young M8 to L5 optical templates (\citealp{2003IAUS..211..163S}; \citealp{2005PASP..117..676R}; \citealp{2006AJ....131.1007B}; \citealp{2008AJ....136.1290R}) to find that the best match are LHS 2924 (a field M9), KPNO-Tau 4 (a young M9.5 BD in the Taurus star forming region; 1~\textendash~10 Myr; \citealp{2002ApJ...580..317B}) and 2MASS~0141-4633 (a young L0$\gamma$ candidate member to the 10~\textendash~40 Myr Tucana-Horologium association; \citealp{2006ApJ...639.1120K}, G2014). The continuum redwards of 8500~\AA\ matches 2MASS~0141-4633 better, however J1207-3900 clearly shows a VO band at 7450~\AA\ which is deeper than that of 2MASS~0141-4633, and similar to that of KPNO-Tau~4. This is consistent with J1207-3900 having a similar spectral type than 2MASS~0141-4633 while being slightly younger. We thus assign it an optical classification of L0$\gamma$ (see Figure~\ref{fig:OPT_comp}).

\subsection{Signs of Low Gravity}\label{sec:lowg}

\begin{figure*}
	\centering
	\subfigure[FeH$_Z$]{\label{fig:FeH_Z}\includegraphics[width=0.49\textwidth]{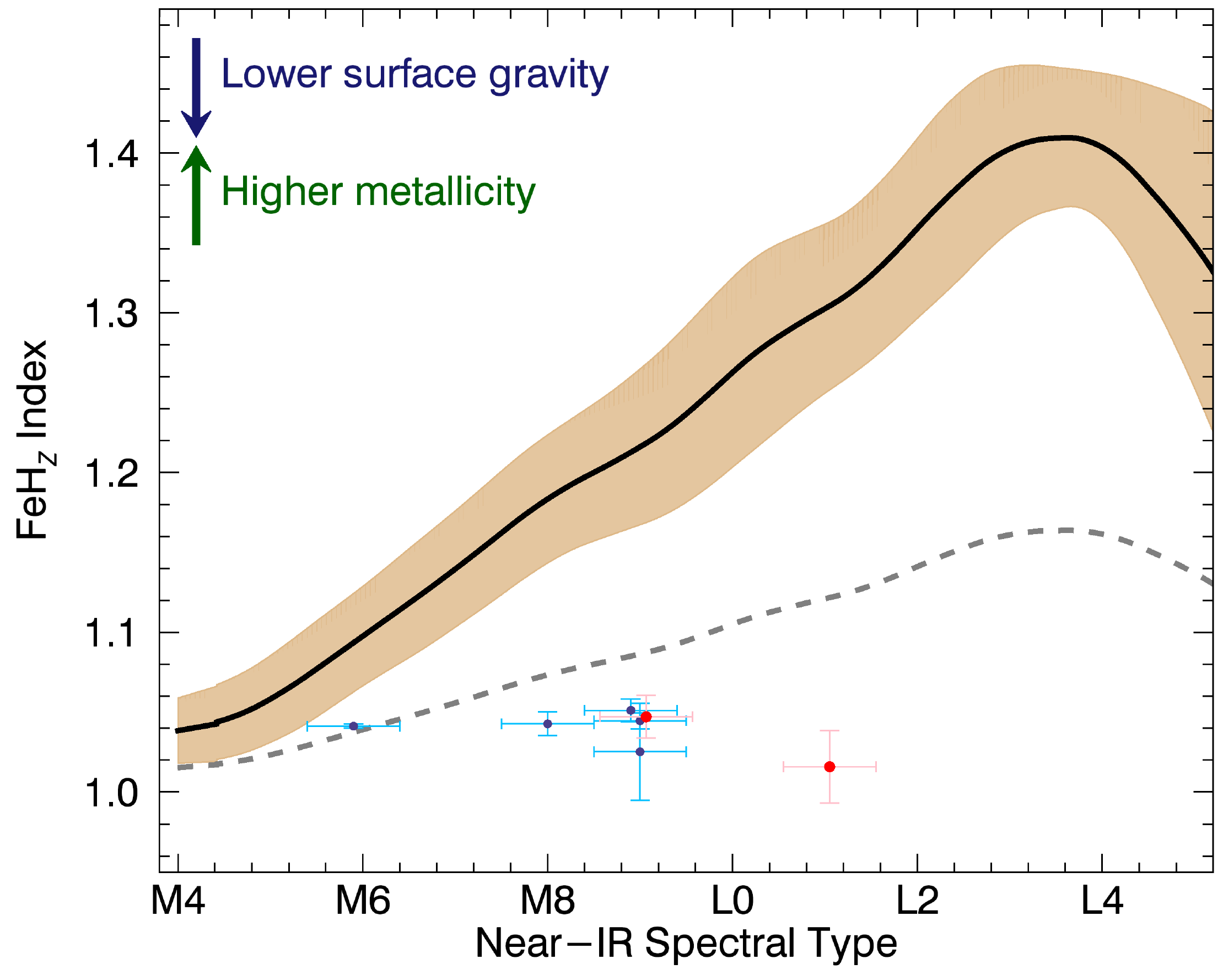}}
	\subfigure[FeH$_J$]{\label{fig:FeH_J}\includegraphics[width=0.49\textwidth]{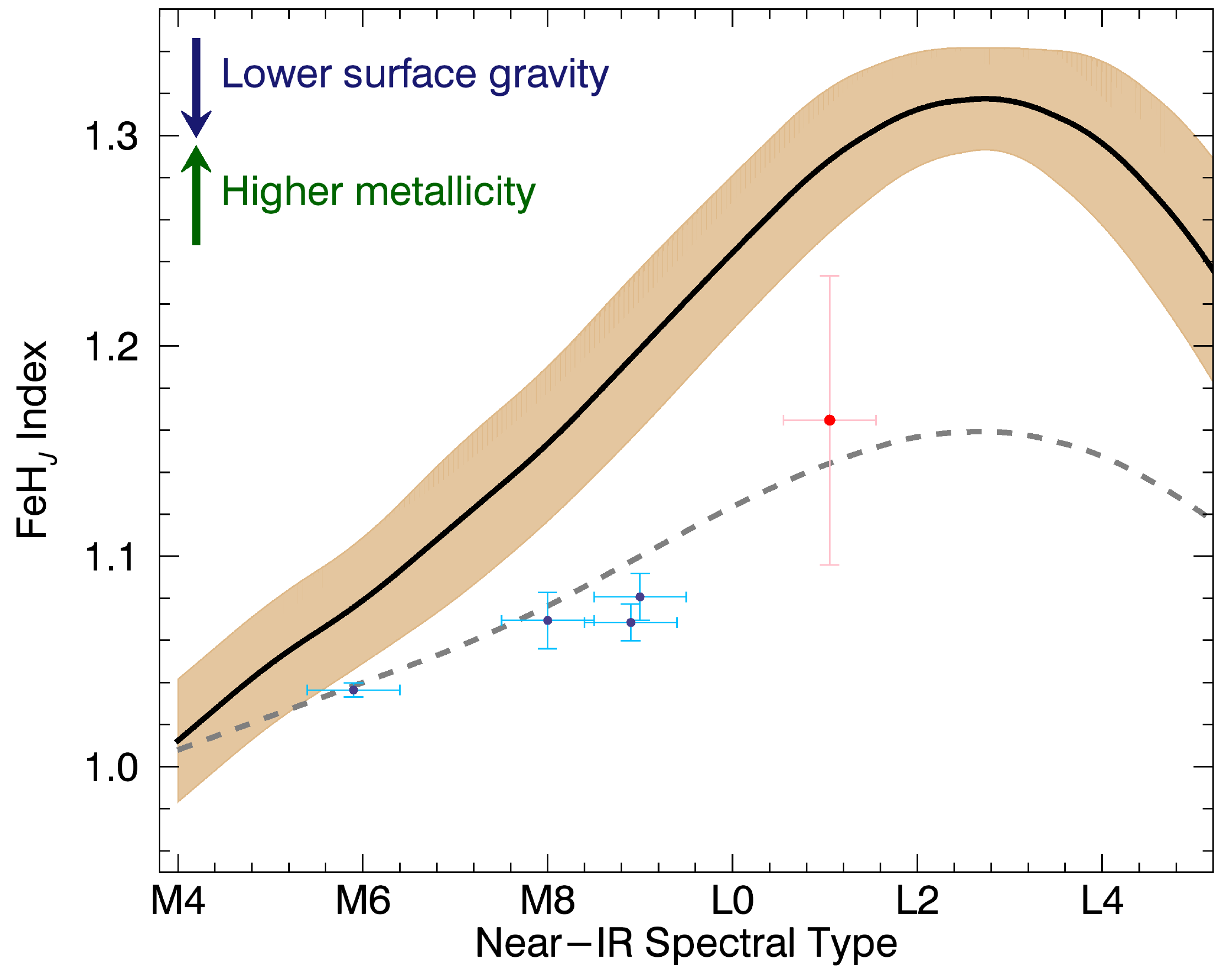}}
	\subfigure[KI$_J$]{\label{fig:KI_J}\includegraphics[width=0.49\textwidth]{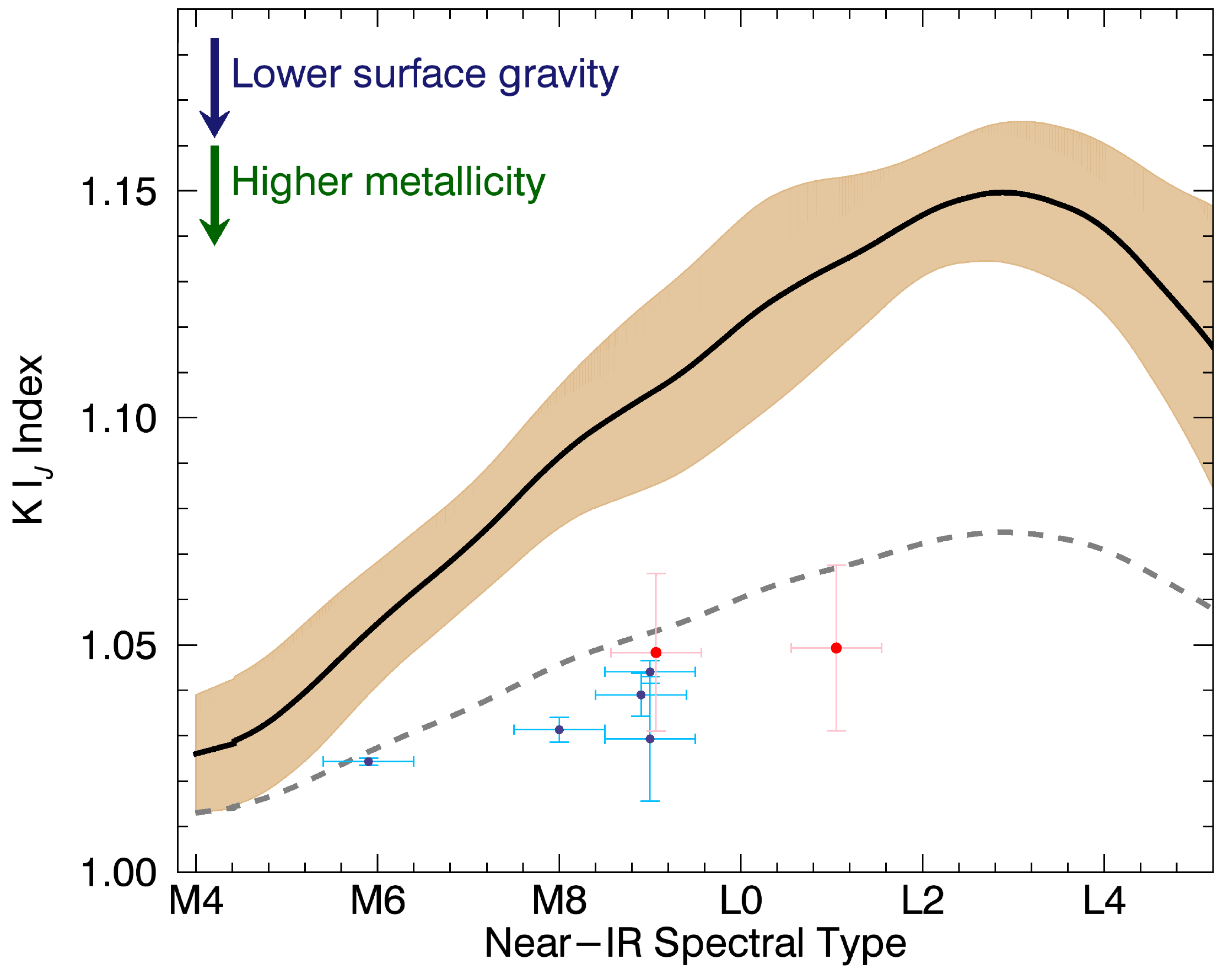}}
	\subfigure[H$_{\mathrm{CONT}}$]{\label{fig:H_CONT}\includegraphics[width=0.49\textwidth]{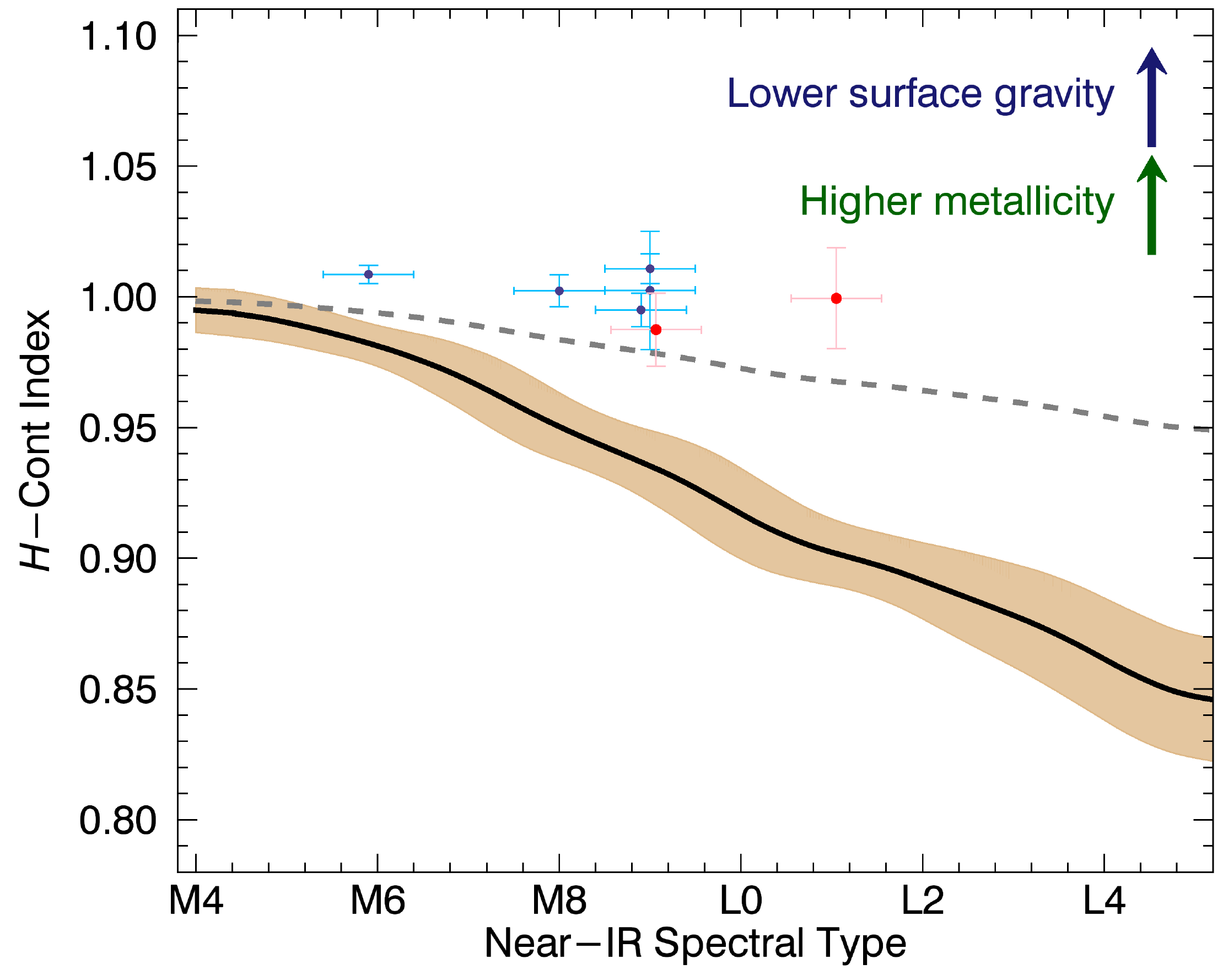}}
	\caption{Spectral indices as defined by \cite{2013ApJ...772...79A} for J1207-3900 and J1247-3816 (red dots), compared to known TWA members (blue dots), the field sequence (thick, black line) and its scatter (beige shaded region). The dotted line represents the delimitation between intermediate (INT-G) gravity and very low gravity (VL-G) regimes. Both candidates have spectral indices consistent with TWA members. Spectral types were offset by small ($<$ 0.15) random subtypes so that vertical error bars can be distinguished. All indices displayed here for J1207-3900 and J1247-3816 were measured using the SpeX prism spectra, except for the FeH$_J$ index which was measured with the cross-dispersed spectrum. The spectra of known TWA members in this figure are those of TWA~22~A (M5) and TWA~34 (M6) obtained respectively from \cite{2009A&A...506..799B} and J. Gagn\'e et al. (in preparation), as well as TWA~27~A (M8), TWA~26 (M9), TWA~28 (M8) and TWA~29 (M9) obtained from \cite{2013ApJ...772...79A}.}
	\label{fig:indices1}
\end{figure*}

We used the NIR gravity classification scheme described in \citeauthor{2013ApJ...772...79A} (\citeyear{2013ApJ...772...79A} ; based on of the gravity-sensitive equivalent widths of \ion{K}{1}, \ion{Na}{1} and continuum features) to analyze the NIR spectra presented here and find that, based on a comparison to other objects of the same spectral types, both objects are clearly VL-G objects (see Figure~\ref{fig:indices1}), as was expected from the visual comparison with low gravity dwarfs. The MAGE spectrum of J1207-3900 was subsequently used in deriving various gravity-dependent indices described in \citeauthor{2009AJ....137.3345C} (\citeyear{2009AJ....137.3345C}; e.g. K-a, K-b, Na-a, Na-b), which also point towards a low surface gravity. We used the IRSA dust extinction tool\footnote[3]{available at \url{http://irsa.ipac.caltech.edu/applications/DUST/}} to verify that both objects are not significantly reddened by interstellar dust along the line of sight, which could potentially mimic some signatures of low gravity. We find that J1207-3900 and J1247-3816 respectively lie in regions of the sky where the E($B$-$V$) extinction is low at $0.0687~\pm~0.0035$ mag and $0.0492~\pm~0.0014$ mag, using a 5 arcminutes search radius \citep{2011ApJ...737..103S}.

\subsection{TWA Membership}\label{sec:bayes}

\begin{figure*}
	\begin{center}
		\includegraphics[width=0.99\textwidth]{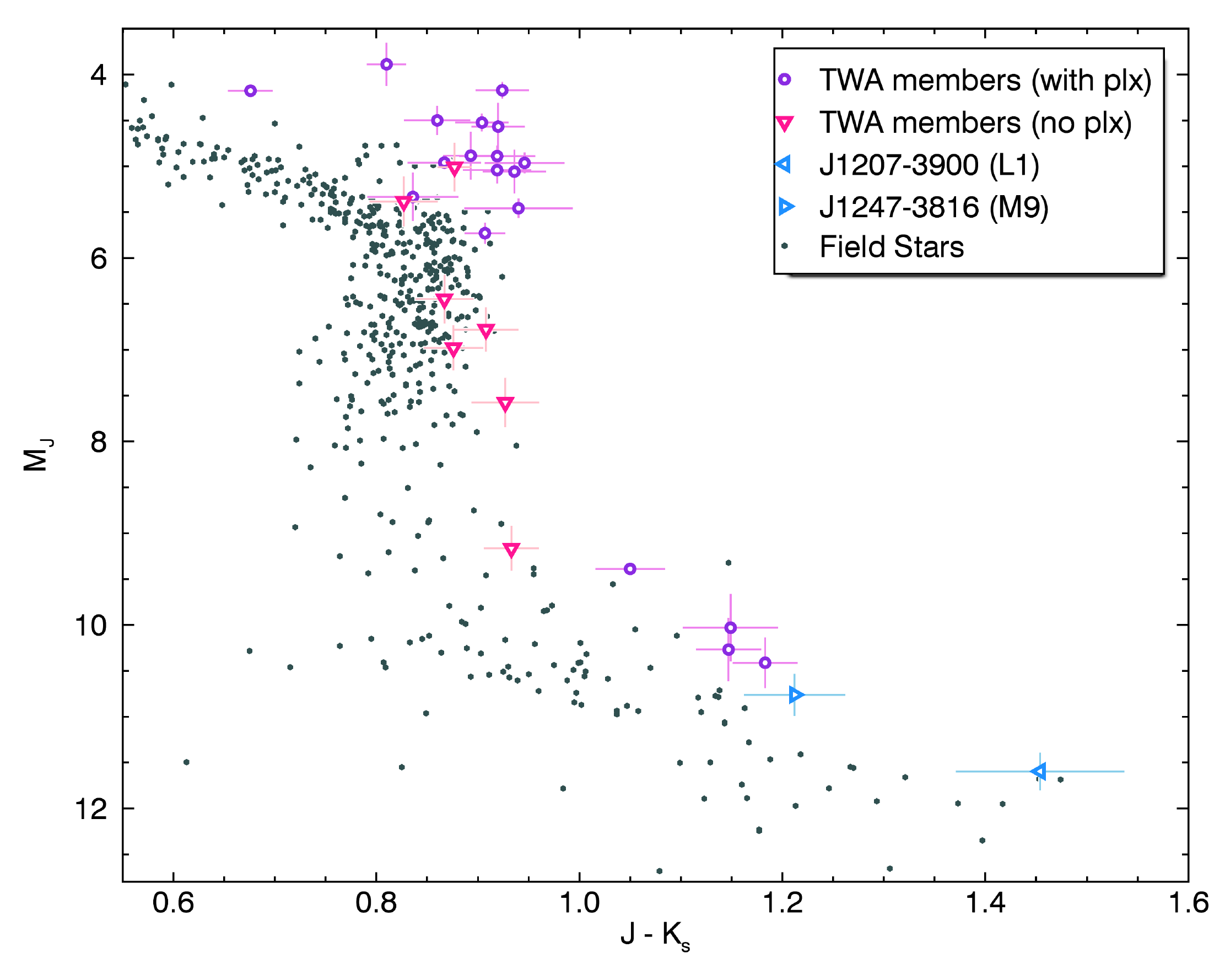}
	\end{center}
	\caption{Color-magnitude sequence for all known primary TWA members and field stars from the CNS3 catalog (\citealp{1991adc..rept.....G}) and Trent Dupuy's \emph{Database of Ultracool Parallaxes} (black dots, \citealp{2012ApJS..201...19D}). We used parallax measurements for TWA candidates when they were available (pink downside triangles ;\citealp{2007ASSL..350.....V} ; \citealp{2008A&A...489..825T} ; \citealp{2013ApJ...762..118W} ; \citealp{Ducourant:2014vn}), or otherwise statistical predictions from BANYAN~II (purple circles). J1247-3816 (right-pointing blue triangle) and J1207-3900 (left-pointing blue triangle) also rely on distance predictions from BANYAN~II, and appear as an extension of the TWA sequence into the L dwarfs regime.}
	\label{fig:PHOT_SEQ}
\end{figure*}

We have applied the BANYAN~II tool described in G2014 to assess the probability that both objects considered here are members of NYAs. We used their sky position, proper motion, spectral types as well as 2MASS and \emph{WISE} photometry as input observables in this analysis, which are then compared to the spatial and kinematic models of each hypothesis considered (TWA, $\beta$~Pictoris, Tucana-Horologium, Columba, Carina, Argus, AB~Doradus and the field) using a naive Bayesian classifier. The spatial and kinematic models are built by fitting the spatial \emph{XYZ} and \emph{UVW} distribution of known bona fide members or synthetic objects from the Besan\c{c}on Galactic model (A. C. Robin et al., in preparataion ; \citealp{2012A&A...538A.106R}) with 3D ellipsoids that are free to rotate along any axes. Following our conclusion that both systems are low gravity dwarfs, we have assumed conservatively that they are younger than 1~Gyr in the construction of the field hypothesis. Using this tool, we find that J1207-3900 and J1247-3816 are both candidate members to TWA with Bayesian probabilities of $99.7$\% and $19.9$\%, respectively. In an ideal case where quantities input in BANYAN~II are strictly independent, the Bayesian probability should represent the best estimate of the probability that a given star be a member of a given NYA, taking into account all available evidence (i.e. data input in the Bayesian inference tool). However, as described in G2014, the Bayesian probabilities determined this way are biased when quantities fed to BANYAN~II (which consists of a \emph{naive Bayesian classifier}) are not strictly independent, which is generally the case in our analysis. A Monte Carlo analysis was thus performed to estimate the unbiased probability that a given object is a field contaminant based on its Bayesian probability. Here, we applied this analysis and found very low field contamination probabilities of 0.004\% and 0.006\%, respectively (meaning that our present Bayesian probabilities are pessimistic). In Table~\ref{tab} we show the radial velocities $v_s$ and distances $d_s$ predicted by Banyan~II, according to the hypotheses that they are actual members to TWA or the field. These estimates were shown by G2014 to be accurate to 8~\% and 1.6~\kms, respectively, when membership is confirmed. \\
We have compared our results to those of BANYAN~I (without using photometry as an observable, see \citealp{2013ApJ...762...88M}), as well as the convergent point analysis (CPA, see \citealp{2013ApJ...774..101R}). In the case of J1207-3900, BANYAN~I yields a membership probability of $99.94$\% for TWA and $0.06$\% for the field with predictions [$v_s = 9.78~\pm~2.2$~\kms, $d_s = 54.0~\pm~5.6$~pc] for TWA, whereas the CPA yields a $91.6$\% probability for TWA, $100.0$\% probability for $\beta$~Pictoris and $81.1$\% probability for Columba with respective predictions of [$v_s = 6.8$~\kms, $d_s = 69.6$~pc], [$v_s = 5.7$~\kms, $d_s = 67.7$~pc] and [$v_s = 12.7$~\kms, $d_s = 79.3$~pc]. Probabilities from BANYAN~I are generally higher than those of BANYAN~II because prior probabilities were set to unity in their analysis, whereas prior probabilities in BANYAN~II are smaller to reflect the smaller populations of NYAs compared to that of the field. Furthermore, we stress the fact that even BANYAN~II probabilities as low as $P_{H_k}~\sim~20$\% for any NYA must be considered as potentially significant, since such values are often found for several known bona fide NYA members that lie 1--2.5$\sigma$ away from the spatial and kinematic locus of bona fide members (see G2014). Probabilities yielded by the CPA are determined individually for each NYA, which means that the total probability can be larger than $100$\%. Thus, both versions of BANYAN agree very well, but the CPA would place J1207-3900 as an ambiguous candidate between $\beta$~Pictoris, TWA and Columba. We do not consider that J1207-3900 is a viable candidate to $\beta$ Pictoris or Columba, since it has been shown by G2014 that such cross-contamination from those two associations to TWA are lower than 3\%, even for low probability TWA candidates. The CPA tool does not consider spatial information or the magnitude of proper motion, and thus often cannot differentiate between a few NYA hypotheses without a radial velocity measurement, especially when their convergent points are close one to another on the celestial sphere, which is the case for TWA, $\beta$ Pictoris and Columba. \\
In the case of J1247-3816, BANYAN~I yields a membership probability of $87.59$\% for TWA and $12.41$\% for the field, with predictions of [$v_s = 7.25~\pm~2.57$~\kms, $d_s = 55.5~\pm~5.9$~pc] for TWA, and the CPA yields $99.3$\% for TWA, $98.7$\% for $\beta$~Pictoris and $96.0$\% for Columba, with respective predictions of [$v_s = 4.0$~\kms, $d_s = 120.7$~pc], [$v_s = 3.1$~\kms, $d_s = 117.0$~pc] and [$v_s = 9.5$~\kms, $d_s = 141.0$~pc]. We note that the relatively smaller probabilities yielded by BANYAN as well as the very large predicted distances from the CPA can both be seen as a consequence of the fact that J1247-3816 has a slightly deviant proper motion compared to TWA members at this sky position. Effectively, the most probable scenario yielded by BANYAN~II places this object at a Galactic position and space velocity $XYZUVW$ of respectively : 30.9~$\pm$~3.1~pc, -49.1~$\pm$~4.9~pc, 26.5~$\pm$~2.7~pc, -2.7~$\pm$~2.2~\kms, -14.5~$\pm$~2.1~\kms\ and -0.7~$\pm$~2.7~\kms, at respectively 1.5$\sigma$ and 3.0$\sigma$ from the spatial and kinematic models used in BANYAN~II. The CPA tool, which is purely kinematic, places J1247-3816 at a larger distance so that it ends up with kinematics closer to those of TWA (at 0.7$\sigma$ of the same kinematic model). \\
\cite{2012ApJ...754...39S} point out that the Lower-Centaurus-Crux (LCC) complex is a possible source of contamination for TWA, however it is located at $\sim$~120~pc (further than typical TWA members at $\sim$~50~pc). Most probable distances derived from BANYAN II for both the TWA and field hypotheses place place J1207-3900 and J1247-3816 at distances of $\sim$~60~pc either for TWA or the field hypothesis, which is not compatible with them being at such a large distance. We conclude that J1207-3900 should be considered as the first isolated L dwarf candidate member to TWA, whereas the membership of J1247-3816 is more ambiguous. Radial velocity and parallax would further constrain their memberships.\\
We have compared these same distance estimates yielded by BANYAN~II as well as 2MASS and \emph{WISE} NIR photometry to the AMES-COND isochrones \citep{2003A&A...402..701B} in combination with CIFIST2011 BT-SETTL atmosphere models (\citealp{2013MSAIS..24..128A} ; \citealp{2013A&A...556A..15R}) in a likelihood analysis, while assuming the age of TWA (8~\textendash~12~Myr), to estimate the masses of both components. We find that J1207-3900 is thus a candidate 11~\textendash~13~\Mjup BD and J1247-3816 is a candidate 14~\textendash~15~\Mjup BD. \\
Both new candidates presented here bring the opportunity of extending the population of TWA members redward in a color-magnitude diagram, up into the L dwarfs regime. In Figure~\ref{fig:PHOT_SEQ}, we show a NIR color-magnitude diagram comparing current TWA members in the literature to field stars and new TWA candidates. In the cases where a trigonometric distance is not available (e.g. for both candidates presented here), we used BANYAN~II to produce a statistical distance estimate corresponding to the kinematics of TWA
and used it to compute a statistical absolute magnitude. It can be seen the TWA sequence is shifted towards redder colors (for late-type objects) and brighter absolute magnitudes (for early-type objects) compared to the field sequence, which is consistent with current evidence on the atmospheric properties of young systems (\citealp{2012ApJ...752...56F} ; \citealp{2013AN....334...85L}). J1207-3900 and J1247-3816 are similarly redder to the field dwarfs sequence, and extend the TWA sequence.

\section{Concluding Remarks}\label{sec:Ldwarf}

The two new candidates to TWA presented here were discovered as part of BASS, an all-sky survey for late-type low-mass stars (LMSs) and BDs in NYAs based on the 2MASS and \emph{WISE} catalogs. This survey has already identified other young objects such as 2MASS~J01033563-5515561 ABb (see \citealp{2013A&A...553L...5D}), and several hundreds of $>$ M5 candidates identified in the same way are currently being followed and results will be published in an upcoming paper (see \citealp{2013arXiv1307.1127G} for more information).

\acknowledgments
We thank our anonymous referee for a thorough and useful revision of this work. We thank Katelyn Allers and Micka\"el Bonnefoy for sharing data and David Rodriguez as well as Michael Liu for useful discussions. This work was supported in part through grants from the Fond de Recherche Qu\'eb\'ecois \textendash  Nature et Technologie and the Natural Science and Engineering Research Council of Canada. This research has benefitted from the SpeX Prism Spectral Libraries, maintained by Adam Burgasser at \url{http://pono.ucsd.edu/\textasciitilde adam/browndwarfs/spexprism}, and the \emph{Database of Ultracool Parallaxes} at \url{http://www.cfa.harvard.edu/\textasciitilde tdupuy/plx/Database\_of\_\\ Ultracool\_Parallaxes.html}. This research made use of ; the SIMBAD database and VizieR catalogue access tool, operated at Centre de Donn\'ees astronomiques de Strasbourg, France \citep{2000A&AS..143...23O} ; data products from the Two Micron All Sky Survey, which is a joint project of the University of Massachusetts and the Infrared Processing and Analysis Center (IPAC)/California Institute of Technology (Caltech), funded by the National Aeronautics and Space Administration (NASA) and the National Science Foundation \citep{2006AJ....131.1163S} ; data products from the Wide-field Infrared Survey Explorer, which is a joint project of the University of California, Los Angeles, and the Jet Propulsion Laboratory (JPL)/Caltech, funded by NASA \citep{2010AJ....140.1868W} ; the NASA/IPAC Infrared Science Archive, which is operated by the JPL, Caltech, under contract with NASA ; the Infrared Telescope Facility, which is operated by the University of Hawaii under Cooperative Agreement NNX-08AE38A with NASA, Science Mission Directorate, Planetary Astronomy Program. This paper includes data gathered with the 6.5 meter Magellan Telescopes located at Las Campanas Observatory, Chile (CNTAC program CN2013A-135).\\

\mbox{~}

\end{document}